\documentclass[]{aa}  

\usepackage{graphicx}
\usepackage{txfonts}
\usepackage{indentfirst}
\usepackage{natbib}
\usepackage{textcomp}
\usepackage{ulem}
\bibpunct{(}{)}{;}{a}{}{,} 
\usepackage[scriptsize]{subfigure}
\usepackage{amsmath}

\usepackage{color}

\begin{document}
  
  \title{A fast, robust, and simple implicit method for adaptive time-stepping on adaptive mesh-refinement grids}
   \author{B. Commer\c con
           \inst{1,2}
           ,
	 V. Debout\inst{3},
 	  \and
            R. Teyssier\inst{4}
          }
   \offprints{B. Commer\c con}

   \institute{ \'Ecole Normale Sup\'erieure de Lyon, CRAL, UMR CNRS 5574, Universit\'e Lyon I, 46 All\'ee d'Italie, 69364 Lyon Cedex 07, France\\
             \email{benoit.commercon@ens-lyon.fr}
    \and Laboratoire de radioastronomie (UMR 8112 CNRS), \'Ecole Normale Sup\'erieure et Observatoire 
de Paris, 24 rue Lhomond, 75231 Paris Cedex 05, France
              \and
              LESIA - Observatoire de Paris, CNRS, UPMC, Universit\'e Paris-Diderot, 5 place Jules Janssen, 92195, Meudon, France
	\and
Universit\"at Z\"urich, Institute f\"ur Theoretische Physik,
Winterthurerstrasse 190, CH-8057 Z\"urich, Switzerland
             }

   \date{Preprint online version: January 6, 2014}

  \abstract  {Implicit solvers present strong limitations when used on supercomputing facilities and in particular for adaptive mesh-refinement codes.}   
  {We present a new method for implicit adaptive time-stepping on adaptive mesh-refinement grids. We implement it in the radiation-hydrodynamics solver we designed for the {\ttfamily RAMSES} code for astrophysical purposes and, more particularly, for protostellar collapse.}
  { We briefly recall the radiation hydrodynamics equations and the adaptive time-stepping methodology used for hydrodynamical solvers. We then introduce the different types of boundary conditions (Dirichlet, Neumann, and Robin) that are used at the interface between levels and present our implementation of the new method in the  {\ttfamily RAMSES} code. The method is tested against classical diffusion and radiation hydrodynamics tests, after which we present an application for protostellar collapse.}
      {We show that using Dirichlet boundary conditions at level interfaces is a good compromise between robustness and accuracy and that it can be used in structure formation calculations. The gain in computational time over our former unique time step method ranges from factors of 5 to 50 depending on the level of adaptive time-stepping and on the problem. We successfully compare the old and new methods for protostellar collapse calculations that involve highly nonlinear physics.}  
       {We have developed a simple but robust method for adaptive time-stepping of implicit scheme on adaptive mesh-refinement grids. It can be applied to a wide variety of physical problems that involve diffusion processes.}   


\titlerunning{}
\authorrunning{B. Commer\c con et al.}
   \maketitle


\section{Introduction}

The study of structure formation in the Universe involves multiscale highly nonlinear physics such as hydrodynamics, radiative transfer, gravity, and magnetic fields. Numerical experiments are the best laboratory for studying these structures, but they remain challenging. Thanks to the  formidable development of supercomputing facilities, these numerical experiments can integrate many different physical processes and use thousands of processors to achieve unprecedented numerical resolution. Nevertheless, efficient scaling often becomes problematic because of the variety of dynamically important physical processes involved. For instance, some physical processes, such as radiative transfer, involve dynamical timescales that are much shorter than in hydrodynamics. If hydrodynamics and radiative transfer are coupled in a unique nonrelativistic system of equations, the time step at which this system can be integrated is limited by the one derived from radiation transport at the speed of light. Implicit methods have thus been developed and coupled to hydrodynamical solvers to handle the short characteristic timescales of physical processes such as the diffusion. In general, hydrodynamical codes use an operator splitting approach with explicit solvers to integrate the Euler equations and implicit solvers to deal with diffusion-like problems. This coupling of explicit and implicit solvers is relatively straightforward and well-studied on uniform grids \citep[e.g.,][]{Turner_stone_01}, but becomes far  more difficult on complex grids, such as those generated by adaptive mesh-refinement \citep[AMR, see the {\ttfamily RAMSES } code][]{teyssier-2002,Commercon_2011}. 

To illustrate the main difficulties of designing an implicit method for AMR grids, let us first consider the simple heat equation in one dimension. The second-order parabolic partial differential equations can be generalized as a diffusion problem following
\begin{equation}
\partial_t U-K\partial^2_{xx}U=0,
\label{heat}
\end{equation}
where $U(x,t)$ is function of position $x$ and time $t$, and $K$ is the diffusion coefficient. Numerically, equation (\ref{heat}) can be integrated with explicit or implicit methods to advance from time level $n$ to time level $n+1$. Explicit discretization of (\ref{heat}) leads to the  Courant Friedrich Levy (CFL) stability condition
\begin{equation}
\Delta t_\mathrm{exp}<\frac{\Delta x^2}{2K},
\label{CFL}
\end{equation}
with $\Delta x$ the size of the discretized mesh. The explicit CFL condition for diffusion equation scales as $\Delta x^2$ and is thus far more stringent than the classical CFL derived for the stability of the hyperbolic system formed by the Euler equations ($\Delta t_\mathrm{hyp}<\Delta x/c_\mathrm{s}$, with $c_\mathrm{s}$ the gas sound speed). This can lead to extensive computing time when both hyperbolic and parabolic equations are treated simultaneously. For that reason, equation (\ref{heat}) is often integrated using an implicit scheme, which is unconditionally stable. The implicit scheme requires solving a matrix system of equation $\mathbb{A}x=b$, where matrix $\mathbb{A}$ has to be inverted to get the vector solution $x$. While matrix inversion does not present strong conceptual difficulties for uniform grids using preconditioned iterative methods, the problem becomes challenging when the grid is complex like the one generated by {\ttfamily RAMSES}. 

Another feature of AMR codes is the use of adaptive time-stepping (ATS) in their hydrodynamical solvers \citep{teyssier-2002,Almgren_2010,Enzo_2013} to speed calculations up. For implicit solvers, ATS is not as straightforward as for an explicit scheme, since an operator integrated with an implicit scheme can affect all the grids of the computational domain.
 Some authors have designed the ATS method for implicit schemes derived from diffusion equations \citep{Howell_Greenough_03,Zhang_et_al_2011}. In these methods, the diffusion equation is updated on a level-by-level basis, and the total radiative energy is conserved by storing flux at the level interfaces.

In a previous paper, \cite{Commercon_2011}, hereafter Paper I,  proposed a method that integrates a diffusion-like equation in the {\ttfamily RAMSES~}\rm code for radiation-hydrodynamics using a two-temperature approach.  The method in Paper I uses a unique time step for all the levels and does not take advantage of the ATS method developed for the hydrodynamical solver in  {\ttfamily RAMSES}\rm. 
The purpose of this paper is to present a new ATS method for implicit solvers on AMR grids in order to speed up the solver of Paper I. We seek to keep the method as simple as possible to allow quick implementation in other codes using ATS for their hydrodynamical solver. 

The paper is organized as follows.  In Sect. 2, we recall the radiation hydrodynamics (RHD) equations we use and briefly present the flux-limited diffusion solver we designed in Paper I. The new implicit solver for the {\ttfamily RAMSES} code is presented in Sect. 3. In Sect. 4, the method is then tested against well-known test cases for diffusion and RHD. As a final test, RHD dense core collapse calculations with very high resolution are performed in Sect. 5. Section 6 summarizes our work and the main results, and presents our perspectives. 

\section{Basic concepts}

\begin{figure*}[thb]
  \centering
  \includegraphics[width=8.cm,height=4cm]{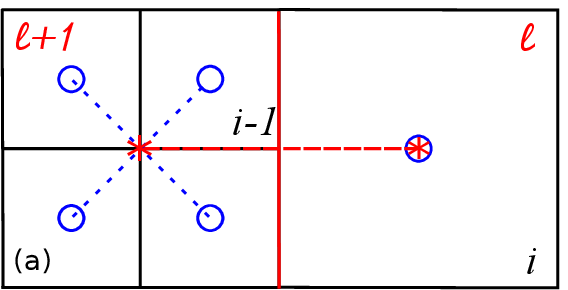}
  \hspace{20pt}
  \includegraphics[width=8cm,height=4cm]{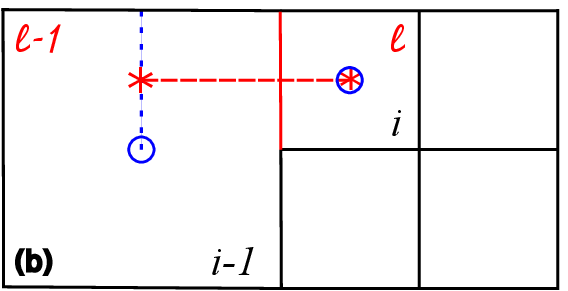}
  \caption{Two dimensional sketch of possible AMR grid configurations at the interface between two levels. {\it Left}: configuration 1 where  cell $i$ is at level $\ell$ and cell $i-1$ at level $\ell+1$ {\it Right}: configuration 2 where cell $i$ is at level $\ell$ and cell $i-1$ at level $\ell-1$. In both sketches, the vertical red line represents the surface over which energy is exchanged when level $\ell$ is updated. Similarly, the horizontal dashed red line represents the gradient over which the flux is computed using the energy marked by a red star (the value is interpolated using the value represented by the attached blue circle).}
\label{config}
\end{figure*}

\subsection{Radiation hydrodynamics in  {\ttfamily RAMSES}}

 In Paper I,  we presented an implementation of a RHD solver into {\ttfamily RAMSES} using the flux-limited diffusion approximation \citep[FLD, e.g.][]{Minerbo_1978JQSRT,Levermore_Pomraning_1981ApJ}. The RHD equations with all the radiative quantities estimated in the comoving frame then read
 \begin{equation}
\left\{
\begin{array}{llll}
\partial_t \rho + \nabla \left[\rho\textbf{u} \right] & = & 0 \\
\partial_t \rho \textbf{u} + \nabla \left[\rho \textbf{u}\otimes \textbf{u} + P \mathbb{I} \right]& =& - \lambda\nabla E_\mathrm{r} \\
\partial_t E_\mathrm{T} + \nabla \left[\textbf{u}\left( E_\mathrm{T} + P_\mathrm{} \right)\right] &= & - \mathbb{P}_\mathrm{r}\nabla:\textbf{u}  - \lambda \textbf{u} \nabla E_\mathrm{r} \\
 & & +  \nabla \cdot\left(\frac{c\lambda}{\rho \kappa_\mathrm{R}} \nabla E_\mathrm{r}\right) \\
\partial_t E_\mathrm{r} + \nabla \left[\textbf{u}E_\mathrm{r}\right]
&=& 
- \mathbb{P}_\mathrm{r}\nabla:\textbf{u}  +  \nabla \cdot\left(\frac{c\lambda}{\rho \kappa_\mathrm{R}} \nabla E_\mathrm{r}\right) \\
 & &  + \kappa_\mathrm{P}\rho c(a_\mathrm{R}T^4 - E_\mathrm{r}),
\end{array}
\right.
\end{equation}
\noindent where $\rho$ is the material density,$\textbf{u}$ is the velocity, $P$ is the thermal pressure, $\kappa_
\mathrm{R}$ is the Rosseland mean opacity,  $\lambda$ is the radiative flux limiter, $E_\mathrm{T}$ is the total energy $E_\mathrm{T}=\rho\epsilon +1/2\rho u^2 + E_\mathrm{r}$ ($\epsilon$ is the internal specific energy), $\kappa_\mathrm{P}$ is the Planck opacity, $E_\mathrm{r}$ is the radiative energy, and $\mathbb{P}_\mathrm{r}$ is the radiation pressure. 
 
 The method presented in paper I is based on an operator splitting scheme, where the hydrodynamical part is integrated using the hyperbolic explicit solver of  {\ttfamily RAMSES} and the radiative energy diffusion and coupling between matter and radiation terms are integrated using an implicit scheme. The method uses a conjugate gradient algorithm in which all the levels of the AMR grid are coupled so that calculations advance in time following the CFL conditions of the finest level of refinement. The main limitation of this method is that it uses a unique time step, and calculations can become very expensive in numerical experiments involving a large hierarchy of AMR levels. In the following, we present an implementation of ATS for the implicit method presented in Paper I.

\subsection{Adaptive time-stepping for hydrodynamics}

Most astrophysical problems  deal with a large range of physical scales, as for instance in star formation, where scales for the size of the cloud (unit of a parsec) and of the protostar (unit of solar radius, R$_\odot\sim10^{-8}$ pc) have to be considered simultaneously. This produces a high level of hierarchy in AMR grids, each level $\ell$ having a size of $\Delta x^\ell$.
For the classical Euler system of equations (conservation of mass, momentum and total energy), a stability condition can be calculated for each level following the CFL condition, namely,
\begin{equation}
\Delta t=C\frac{\Delta x}{|\textbf{u}|+c_\mathrm{s}},
\end{equation}
where $C<1$ is the CFL number and $|\textbf{u}|$ is the fluid velocity norm.

To integrate the Euler equations on AMR grids, a unique time step can be used, meaning that all the levels evolve with the same time step, given by the CFL condition on the finest level $\ell_\mathrm{max}$. This method is very powerful, but can be expensive when a high number of AMR levels is used. In the ATS scheme, each level evolves with its own time step which considerably reduces the CPU time. {\ttfamily RAMSES} uses ATS for the hydrodynamics solver \citep{teyssier-2002}, following the rule that the time step of a level $\ell$ equals always the sum of two  time steps on the finer level, i.e., $\Delta t^\ell=\Delta t^{\ell+1}_1+\Delta t^{\ell+1}_2$. Ideally, if all the levels of the grid hierarchy are effective and if the problem is isothermal and has a uniform velocity, we have $\Delta t^{\ell_\mathrm{max}}=2^{\ell_\mathrm{max}}\Delta t^{0}$, where $\Delta t^{0}$ is the time step of the coarser level. The scheme is only first-order accurate at level interfaces, but the errors are localized only at level interfaces and generally the ratio between the interface surface and the volume of the computational domain is relatively small \citep[otherwise a uniform grid would be used,][]{teyssier-2002}.  In addition, truncation errors can propagate only if waves move from coarse to fine grids, which is not the case in accretion flows (the accretion shock moves from fine to coarse). In the {\ttfamily RAMSES} implementation, the fine levels are updated first. When a cell $i+1$ at level $\ell$ is updated, the flux $F^{n+\Delta t^{\ell-1}}_{i+1/2}$ that crosses the interface with a cell $i$ at a coarser level $\ell-1$ during the two fine time steps $\Delta t_1^\ell+\Delta t_2^\ell$ is 
\begin{equation}
F^{n+\Delta t^{\ell-1}}_{i+1/2}=\frac{1}{\Delta t_1^\ell+\Delta t_2^\ell}\left(\Delta t_1^\ell F^{n+\Delta t_1^\ell}_{i+1/2} + \Delta t_2^\ell F^{n+\Delta t_1^\ell+\Delta t_2^\ell}_{i+1/2}\right),\label{mean_flux}
\end{equation}
such that the total conservative variables (mass, momentum and total energy) are conserved. The aim of this paper is to couple an implicit solver to the hydrodynamical ATS scheme used in {\ttfamily RAMSES}.

\section{Numerical method}

\subsection{Definitions}
Let us consider the following diffusion equation on the radiative energy $E_\mathrm{r}$
\begin{equation}
\frac{\partial E_\mathrm{r}}{\partial t}= \nabla. K\nabla E_\mathrm{r},
\label{Er_diff}
\end{equation}
The finite volume discretization in the $x$ direction\footnote{We assume that the radiative energy is uniform in the $y$ and $z$ direction for a given position on the $x$ axis, i.e., plane parallel approximation.} of equation (\ref{Er_diff})  using an implicit scheme gives
\begin{eqnarray}
\frac{E_{\mathrm{r},i}^{n+1}-E_{\mathrm{r},i}^{n}}{\Delta t}V_i & = &   K_{i+1/2}\frac{E_{\mathrm{r},i+1}^{n+1}-E_{\mathrm{r},i}^{n+1}}{\Delta x}S_{i+1/2} \nonumber\\
 & & -  K_{i-1/2}\frac{E_{\mathrm{r},i}^{n+1}-E_{\mathrm{r},i-1}^{n+1}}{\Delta x}S_{i-1/2},\label{Er_imp}
\end{eqnarray}
where $V_i$ is the volume of cell $i$, $S_{i\pm1/2}$ the surface of exchange  between cell $i$ and cell $i\pm1$, $K_{i\pm1/2}$ is the mean diffusion coefficient computed at the cell interface (e.g., $K_{i\pm1/2}=(K_{i\pm1}+K_i)/2$). Equation (\ref{Er_imp}) can be written in the form
\begin{equation}
-C_{i-1/2}E_{\mathrm{r},i-1}^{n+1} +(1 + C_{i-1/2} + C_{i+1/2})E_{\mathrm{r},i}^{n+1} -C_{i+1/2} E_{\mathrm{r},i+1}^{n+1} = 
E_{\mathrm{r},i}^{n},
\label{Er_imp1}
\end{equation}
where $C_{i\pm1/2}=K_{i\pm1/2}S_{i\pm1/2}\Delta t /(\Delta x V_i)$. Equation (\ref{Er_imp1}) forms a matrix system, $\mathbb{A}x=b$, where matrix $\mathbb{A}$ has to be inverted to get the new value of the radiative energy $E_{\mathrm{r}}^{n+1}$ (the solution vector $x$).
The $C_{i\pm1/2}$ coefficient depends on the grid configuration, namely if the neighboring cells $i-1$ and $i+1$ are or not at the same level of refinement as cell $i$. In the first configuration, fig. \ref{config}a, cell $i$ is at level $\ell$ and cell $i-1$ at level $\ell+1$. In the simplest case, the neighboring cells at level $\ell+1$ are interpolated on a coarser cell at level $\ell$, so that the $C^{\ell\rightarrow\ell+1}$ coefficient calculations is reduced to the one on a uniform mesh. We thus have
\begin{eqnarray}
C^{\ell\rightarrow\ell+1}_{i-1/2} & = & \frac{K_{i-1/2}(\Delta x^{\ell})^{ndim-1}\Delta t^\ell}{\Delta x^{\ell}(\Delta x^{\ell})^{ndim}},\nonumber\\
  & = & \frac{K_{i-1/2}\Delta t^\ell}{(\Delta x^{\ell})^2},
\end{eqnarray}
where $ndim$ is the number of dimensions of the problem. In the opposite case, fig. \ref{config}b, where cell $i$ is at level $\ell$ and cell $i-1$ at level $\ell-1$, we have
\begin{eqnarray}
C^{\ell\rightarrow\ell-1}_{i-1/2} & = & \frac{K_{i-1/2}(\Delta x^{\ell})^{ndim-1}\Delta t^\ell}{\frac{3\Delta x^{\ell}}{2}(\Delta x^{\ell})^{ndim}}, \nonumber\\
  & = & \frac{2}{3}\frac{K_{i-1/2}\Delta t^\ell}{(\Delta x^{\ell})^2}.
 \end{eqnarray}
We define $\tilde{K}$ as the  mean diffusion coefficient at cells interface, and $A^\ell=\tilde{K}\Delta t^\ell/(\Delta x^\ell)^2$. For a given level $\ell$, we thus have three types of coefficient $C$, namely  $C^{\ell\rightarrow\ell+1}$, $C^{\ell\rightarrow\ell}$, and, $C^{\ell\rightarrow\ell-1}$, depending on the grid configuration at cells interface
\begin{eqnarray}
C^{\ell\rightarrow\ell-1} & = &\frac{2}{3}A^\ell,\\
C^{\ell\rightarrow\ell\hspace{8pt}} & = &A^\ell,\\
C^{\ell\rightarrow\ell+1} & = & A^\ell.
\end{eqnarray}
In the previous implementation of Paper I, the cells at level $\ell+1$ are not interpolated at level $\ell$ and the $C^{\ell\rightarrow\ell+1}$ coefficient equals $2^{3-ndim}A^\ell/3$.

In the following, we solve the diffusion equation for each level $\ell$ independently from the other levels. In the case where a cell of level $\ell$ is at an interface with a coarser or a finer level, we need to specify a boundary condition at the level interfaces to solve the matrix system given by equation (\ref{Er_imp1}). We now study different types of boundary condition at level interfaces that can be used to compute the corresponding flux between cells $i$ and $i-1$, so that $C_{i-1/2}=C^{\ell\rightarrow\ell+1}$ or $C_{i-1/2}=C^{\ell\rightarrow\ell-1}$. If cell $i$ is at level $\ell$, cell $i-1$ can be either at level $\ell+1$ or at level $\ell-1$. We assume that cells $i$ and $i+1$ are at the same level $\ell$. 

\subsection{Different types of boundary conditions}

\subsubsection{Dirichlet boundary condition}
The Dirichlet boundary is an imposed boundary condition, i.e., $E^{n+1}_{\mathrm{r},i-1} = E^{n}_{\mathrm{r},i-1}$. 
Equation (\ref{Er_imp1}) then reads 
\begin{equation}
(1 + C_{i-1/2} + C_{i+1/2})E_{\mathrm{r},i}^{n+1} -C_{i+1/2} E_{\mathrm{r},i+1}^{n+1} = 
E_{\mathrm{r},i}^{n} + C_{i-1/2}E_{\mathrm{r},i-1}^{n},\nonumber
\end{equation}
where we moved the terms corresponding to $E_{\mathrm{r},i-1}$ on the rand-hand-side (RHS) of the matrix system, i.e., in the $b$ vector.

\subsubsection{Neumann boundary condition}

The Neumann boundary corresponds to an imposed flux $F_{i-1/2}$, i.e., $C_{i-1/2}(E^{n+1}_{\mathrm{r},i}-E^{n+1}_{\mathrm{r},i-1}) = F_{i-1/2}$.
Equation (\ref{Er_imp1}) reads 
\begin{equation}
(1 + C_{i+1/2})E_{\mathrm{r},i}^{n+1} - C_{i+1/2} E_{\mathrm{r},i+1}^{n+1} = 
E_{\mathrm{r},i}^{n} + F_{i-1/2},\nonumber
\end{equation}
where the imposed flux is also computed in the RHS of the matrix system.

\subsubsection{Robin boundary condition}
The Robin boundary corresponds to a mix between the Dirichlet and the Neumann boundary conditions, i.e., the energy exchange at level interfaces corresponds to $(1-\alpha)F_{i-1/2} +  \alpha (E^{n+1}_{\mathrm{r},i}-E^{n}_{\mathrm{r},i-1})/\Delta x$. Equation (\ref{Er_imp1}) reads 
\begin{eqnarray}
(1 + \alpha C_{i-1/2} + C_{i+1/2})E_{\mathrm{r},i}^{n+1} - C_{i+1/2} E_{\mathrm{r},i+1}^{n+1} & = & 
E_{\mathrm{r},i}^{n} + \alpha C_{i-1/2}E_{\mathrm{r},i-1}^{n}\nonumber\\
 & & + (1-\alpha)F_{i-1/2},\nonumber
\end{eqnarray}
where $\alpha$ is an {\it ad-hoc} parameters, that gives the weight of each type of boundary conditions ($0\leq\alpha\leq1$). Robin boundary conditions  can be not only used as physical boundary conditions, but also as virtual boundary conditions when solving large matrix system with a parallel algorithm on subdomains \citep[e.g., ][]{Tang_1992}.

\subsection{Implementation in \ttfamily{RAMSES}\rm}

We distinguish two types of interface: the coarse-to-fine and the fine-to-coarse. When the neighboring cells are at the same level, no special trick for the flux calculation is required.  As we already mentioned, the fine levels are updated first in the {\ttfamily RAMSES} ATS algorithm. We define $\Delta x$ as the size of the the grid mesh at level $\ell$ (the mesh size is uniform in all direction). In the following, we assume that ATS is used for all levels by default.

\subsubsection{Fine-to-coarse interface}

We consider the case where cell $i-1$ is at level $\ell-1$. In this case, we use a simple Dirichlet boundary condition, where the radiative energy at the coarser level is set to be constant during the two time steps of level $\ell$. To compute the radiative energy gradient at the interface, we also assume that the radiative energy is uniform within the neighboring coarse cell (no interpolation, see fig. \ref{config}b), so that
\begin{equation}
\tilde{F}_{i-1/2} = -K_{i-1/2}\frac{E_{\mathrm{r},i}^{n+1}-E_{\mathrm{r},i-1}^{n}}{\frac{3}{2}\Delta x}.
\end{equation}
The contribution of $E_{\mathrm{r},i-1}^{n}$ is then moved to the RHS of the matrix system. 
 At the end of the diffusion update, the flux $\tilde{F}_{i-1/2}$ is stored at the interface boundary following equation (\ref{mean_flux}), ensuring energy conservation.

\subsubsection{Coarse-to-fine interface}

In this case, we allow the use of the three different types of boundary conditions presented previously. As will be explained below, each has its pros and cons. In the case of the Neumann boundary condition, the flux that is imposed at the level interfaces is given by the flux that has been stored during the update of level $\ell+1$. This ensures energy conservation but can lead to negative energy problems (see \ref{limitation}). For the Dirichlet boundary type, a restriction operation is performed on the neighbor parent cell (oct) using   the updated value of the leaf cells at level $\ell+1$, i.e., $E_{\mathrm{r},i-1}^{n+\Delta t^{\ell+1}}=E_{\mathrm{r},i-1}^{n+\Delta t_1^{\ell}+\Delta t_2^{\ell}}$. The Robin condition uses a mix between the stored flux and the updated neighbor energy, with parameter $\alpha$ being a user defined parameter. $(1-\alpha)$ gives the relative amount of the energy that is conserved at the interface.

\subsubsection{Energy loss with Dirichlet BC at coarse-to-fine interface}

In the case of a Dirichlet boundary condition, the error made on energy conservation at the interface can be computed analytically. We assume that the radiative energy is uniform within the oct at level $\ell +1$, so that the restriction operation on the oct gives $(\tilde{E}_{\mathrm{r},i}^{n+\Delta t^{\ell-1}})^{\ell-1}=(E_{\mathrm{r},i}^{n+\Delta t^{\ell-1}})^\ell$.

The flux that crosses the surface $S^\ell=(\Delta x^\ell)^{ndim-1}$ between cell $i$ and cell $i-1$ during the update of level $\ell$ is given by equation (\ref{mean_flux})
\begin{equation}
F^{\ell\rightarrow\ell-1}_{i-1/2}=\frac{1}{2}\frac{K}{\frac{3}{2}\Delta x^\ell}\left( E_{\mathrm{r},i}^{n+\Delta t^{\ell}}+ E_{\mathrm{r},i}^{n+2\Delta t^{\ell}}-E_{\mathrm{r},i-1}^{n}\right),
\end{equation}
where we assume that $\Delta t_1^{\ell}=\Delta t_2^{\ell}=1/2\Delta t_1^{\ell-1}$ and that the diffusion coefficient $K$ is constant.
Similarly, the flux that crosses the same surface during level $\ell-1$ update equals
\begin{equation}
F^{\ell-1\rightarrow\ell}_{i-1/2}=\frac{K}{2\Delta x^\ell}\left( E_{\mathrm{r},i-1}^{n+2\Delta t^{\ell}}-E_{\mathrm{r},i}^{n+2\Delta t^{\ell}}\right).
\end{equation}

Energy conservation requires that $F^{\ell-1\rightarrow\ell}_{i-1/2}=-F^{\ell\rightarrow\ell-1}_{i-1/2}$. In our implementation, the energy mismatch is
\begin{eqnarray}
\Delta E_\mathrm{r}&=&\frac{K}{6\Delta x^\ell}\left( E_{\mathrm{r},i}^{n+\Delta t^{\ell}} - E_{\mathrm{r},i}^{n+2\Delta t^{\ell}} +E_{\mathrm{r},i}^{n+\Delta t^{\ell}}-E_{\mathrm{r},i-1}^{n}\right) \\ 
& & +\frac{K}{2\Delta x^\ell} \left( E_{\mathrm{r},i-1}^{n+2\Delta t^{\ell}} - E_{\mathrm{r},i-1}^{n}  \right).
\end{eqnarray}
We see that three quantities contribute to the energy loss. First, it is proportional to the rate of change of the energy during the fine level updates ($E_{\mathrm{r},i}^{n+\Delta t^{\ell}} - E_{\mathrm{r},i}^{n+2\Delta t^{\ell}} $) and also directly to the first intermediate flux ($E_{\mathrm{r},i}^{n+\Delta t^{\ell}}-E_{\mathrm{r},i-1}^{n}$). Second, it mainly depends on the energy change on the coarse cell itself during the coarse update ($E_{\mathrm{r},i-1}^{n+2\Delta t^{\ell}} - E_{\mathrm{r},i-1}^{n}$). Energy conservation can then be strongly violated in case of large energy gradients and energy change (early time of an energy  pulse propagation).

\begin{figure} [thb]
\centering
\includegraphics[width=7.cm,height=4cm]{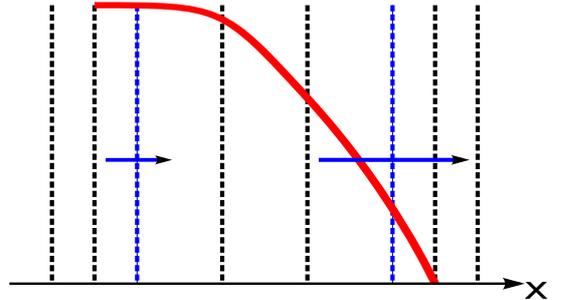}
\caption{Illustration of the negative energy problem. The grid meshes are represented by the dashed lines and the energy gradient by the red curve. The stored flux at the level interfaces (dashed blue) are represented by the blue arrow. In this case, more energy goes out than comes in when updating the coarse level.}
\label{pb}
\end{figure}

\subsubsection{Implicit update}

We follow the same iterative method as in Paper I to solve the implicit system of the coupled equations governing the radiative and gas internal energies using the two-temperature approach. As in Paper I the coupled system of equations is reduced to a single equation on the radiative energy thanks to the linearization of the emission term. The only difference is that the iterative method is called on a level-by-level basis after the hyperbolic update of each AMR level. We use a conjugate gradient algorithm with a diagonal preconditioner. The stopping criterion is based on the L$_2$ norm of the residual $r^{(j)}/r_0<\epsilon_\mathrm{conv}$ (where $r_0$ is the initial residual) and on the L$_\infty$ norm of the relative change of the radiative energy between two iterations $(j)$ and $(j-1)$, i.e., max$|E_\mathrm{r}^{(j)}-E_\mathrm{r}^{(j-1)}|/E_\mathrm{r}^{(j-1)}<\epsilon_\mathrm{conv}$. More details on the iterative solver and the two equations implicit system solver can be found in Paper I.

\begin{figure} [thb]
\centering
\includegraphics[width=9.cm,height=7.65cm]{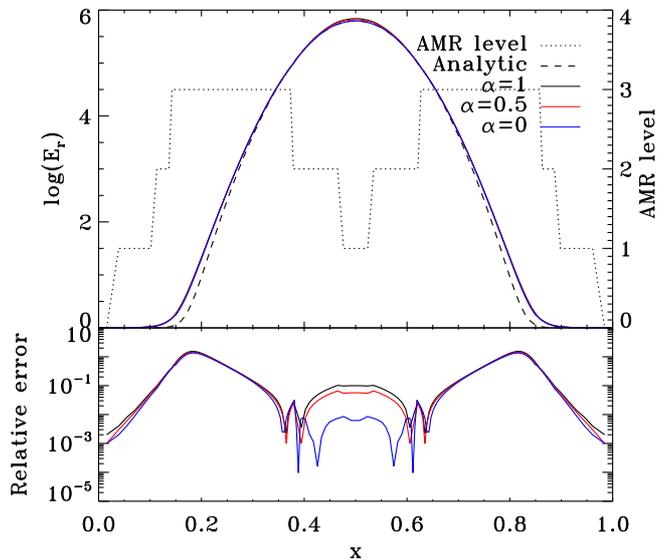}
\caption{Linear diffusion test. {\it Top: } radiative energy profile at time $t=2\times10^{-13}$ for three calculations using $\alpha=1$ (black), $\alpha=0.5$ (red), and $\alpha=0$ (blue), i.e., using, respectively, Dirichlet, Robin, and Neumann boundary conditions at level interfaces. The analytical solution is represented by the dashed line. The right axis shows the AMR levels, i.e., the effective resolution profile (dotted line).  {\it Bottom: } Corresponding relative error as a function of the distance. }
\label{dirac}
\end{figure}

\begin{figure} [thb]
\centering
\includegraphics[width=9.cm,height=7.65cm]{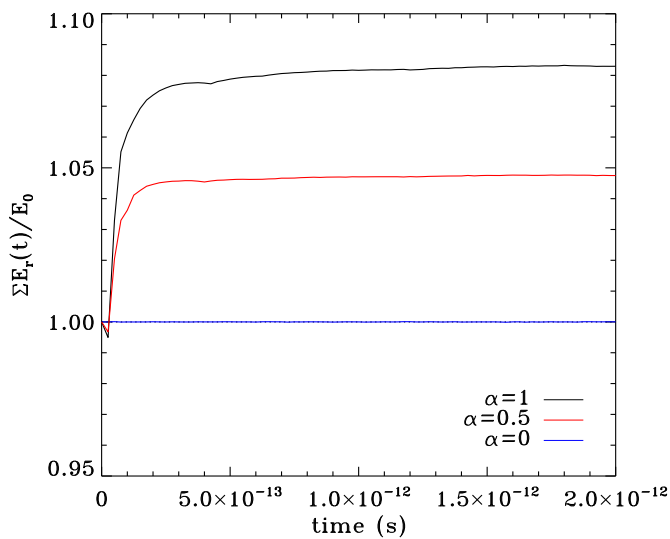}
\caption{Linear diffusion test: energy conservation as a function of time for the three calculations using  $\alpha=1$ (black), $\alpha=0.5$ (red), and $\alpha=0$ (blue). }
\label{cons_nrj}
\end{figure}

\subsubsection{Limitations and comparison to other methods\label{limitation}}

The first limitation of our scheme is that it uses a fully implicit method, which is first-order in time so that it is generally dominated by the truncation error due to the time discretization. This could be improved in the future by using a Crank-Nicolson integrator scheme.

Secondly, Dirichlet boundary condition at the fine-to-coarse interface is a common approximation made by various authors \citep{Howell_Greenough_03,Zhang_et_al_2011}. It is a relatively crude approximation since it can lead to flux over-(under-) estimate at level interfaces. This flux is stored at the end of the fine level update in order to allow for energy conservation when updating the coarse level. Nevertheless, this flux, $\tilde{F}_{i-1/2}$, has been computed using a desynchronized value of the coarse level energy, so that it actually does not correspond to the correct flux $F^{n+1}_{i-1/2}$. Energy conservation is then ensured using artificial flux corrections. This inaccuracy can be improved using a multilevel solver as proposed by \cite{Howell_Greenough_03}. However, for the sake of simplicity, we have decided not to use this in the present work, a choice justified by  the strong performance of our method in the tests below. Another drawback of flux storage at level interfaces is that the location where energy is stored is certainly not the correct one. Contrary to the case of the ATS method for the explicit hyperbolic solver, information can propagate across many cells during a single time step with the implicit scheme. For instance, energy could have been transported much further than the cell boundary to even coarser levels. On the other hand, we showed that using a Dirichlet boundary condition at the coarse-to-fine interface leads to unavoidable loss (or gain) of energy. This can be improved using the Robin boundary condition. 

Sometimes we experienced severe problems using the flux conservation method in the case where a coarse level $\ell-1$ is surrounded by finer level $\ell$ with a gradient of energy following the sketch presented in figure \ref{pb}. The flux stored at the left interface is lower that the one on the right. Using an implicit scheme can lead to large fluxes, and it is straightforward to see that energy conservation can lead to negative energy on level $\ell-1$ when it will be updated. (More energy goes out than comes in.) This is problem dependent, but unavoidable. Using Dirichlet boundary conditions avoids negative energies and is in that sense more robust.

\section{Tests}

In this section, we perform a suite of numerical experiments to test and validate our method. We first test the diffusion operator using AMR, and then we perform a full RHD test. 

\subsection{Linear diffusion test}

This first test is the same as in Paper I, Sec. 4.1. It consists in letting an energy pulse diffuse in a uniform medium. We only consider the radiative energy diffusion operator in the optically thick limit  
\begin{equation}
\frac{\partial E_\mathrm{r}}{\partial t} - \nabla \cdot\left(\frac{c}{3 \rho \kappa_\mathrm{R}} \nabla E_\mathrm{r}\right) = 0.
\end{equation}

We consider a box of length L=1. The initial radiative energy corresponds to a delta function, namely it is equal to 1 everywhere in the box, except at the center where it  equals $E_\mathrm{r,L/2}\Delta x=E_0=10^5$. We choose $\rho\kappa_\mathrm{R}=1$. We apply periodic boundary conditions to ensure energy conservation. We use three levels of refinement with a coarse grid of 32 cells (effective resolution of 256 cells at the maximum level of refinement $\ell=3$). The mesh is refined when the  radiative energy relative gradient exceeds 25\% in a cell (e.g., if in cell $i$ we have $2|E_{\mathrm{r},i}-E_{\mathrm{r},i\pm 1}|/(E_{\mathrm{r},i}+E_{\mathrm{r},i\pm1})>0.25$). Each additional level uses a time step twice as smaller as the coarser one. The coarsest time step is kept constant to $\Delta t^0=3.125\times10^{-15}$ which gives a diffusion  CFL of $\sim 4$ on the maximum level of refinement. We present three calculations using different values of $\alpha$  corresponding to different boundary condition at level interfaces, namely $\alpha=1$ for Dirichlet, $\alpha=0.5$ for Robin, and $\alpha=0$ for Neumann. The convergence criterion $\epsilon_\mathrm{conv}$ is set to $10^{-8}$.

Figure \ref{dirac}, top panel, shows the radiative energy profiles (left axis) for the three calculations using $\alpha=1$ (black), $\alpha=0.5$ (red), and $\alpha=0$ (blue) at time $t=2\times 10^{-13}$, as well as the AMR levels (right axis, dotted line). The three calculations match  the analytic solution (dashed line) remarkably well, even when energy is not conserved at the interfaces. Differences only appear at the center of the domain as illustrated in the bottom panel showing the corresponding relative error profiles. The errors remain of the order of a few percent and below except in the tail of the diffusion patch. At this location, the increase in the relative error is the same in the three models as a consequence of the truncation error due to the time discretization (first-order). Figure \ref{cons_nrj} shows the total energy conservation as a function of time for the three calculations. As expected, energy is perfectly conserved in the case where $\alpha=0$, but increases by a few percent (up to $\sim 8$\%) in the cases where $\alpha \ne 0$. Using $\alpha=0.5$  results in better energy conservation  compared the Dirichlet boundary condition. Finally, it is interesting to note that once the radiative energy  profile becomes smoother ($t>10^{-12}$), the energy gain (or loss) is stabilized, indicating that energy conservation is much improved thereafter. This first test indicates that using Dirichlet boundary condition and ATS is reasonable even for extreme initial conditions such as a Dirac pulse. In the following, all the calculations we present have been run using the Dirichlet approximation when ATS is used.

\subsection{Equilibrium test with nonlinear diffusion coefficient}

\begin{figure} [t]
\centering
\includegraphics[width=9.cm,height=7.65cm]{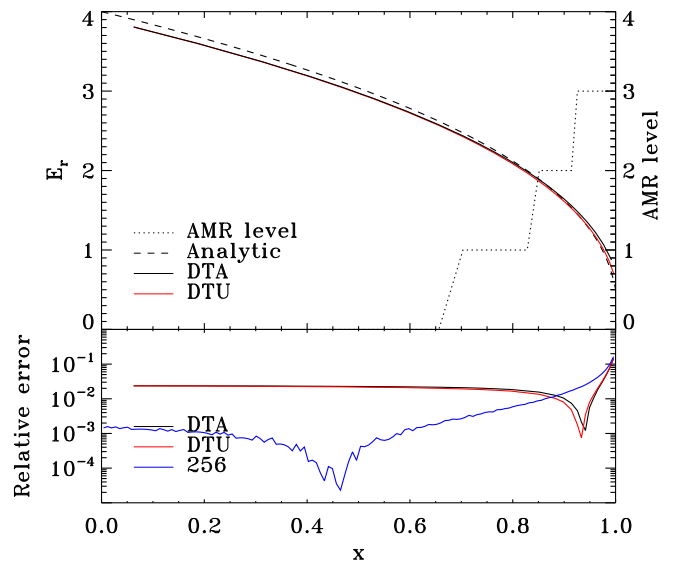}
\caption{Equilibrium test with nonlinear diffusion coefficient. {\it Top:} Radiative energy profile for the DTA (adaptive time step, black) and DTU (unique time step, red) models, and for the stationary analytic solution (dashed line). The right axis indicates the  AMR levels of the DTA and DTU models (dotted line).{\it Bottom:} Relative error profiles for the DTA (black), DTU (red) and uniform grid (256 cells, blue) models.}
\label{equilibrium}
\end{figure}
\begin{figure} [thb]
\centering
\includegraphics[width=9.cm,height=7.65cm]{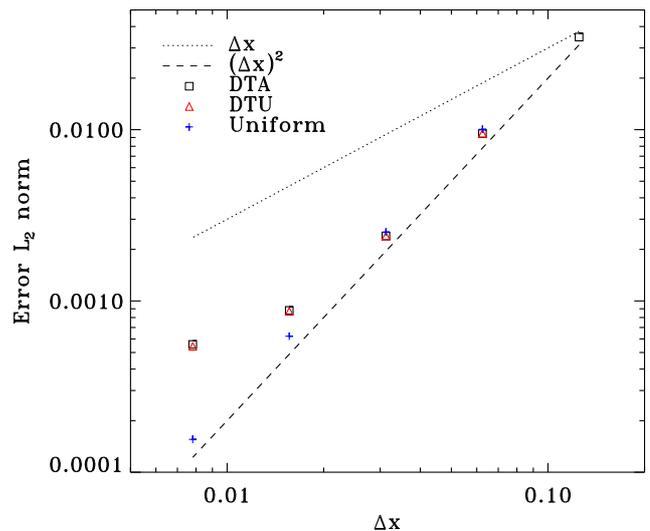}
\caption{L$_2$ norm of the error as a function of the minimum grid size for the three models: DTA (black square), DTU (red triangle), and uniform resolution (blue cross). The dotted line gives the slope that is proportional to $\Delta x$ (first-order accuracy in space) and the dashed line the $(\Delta x)^2$ slope (expected second-order accuracy).}
\label{error_l2}
\end{figure}

\begin{figure*} [t]
\centering
\includegraphics[width=9.cm,height=7.65cm]{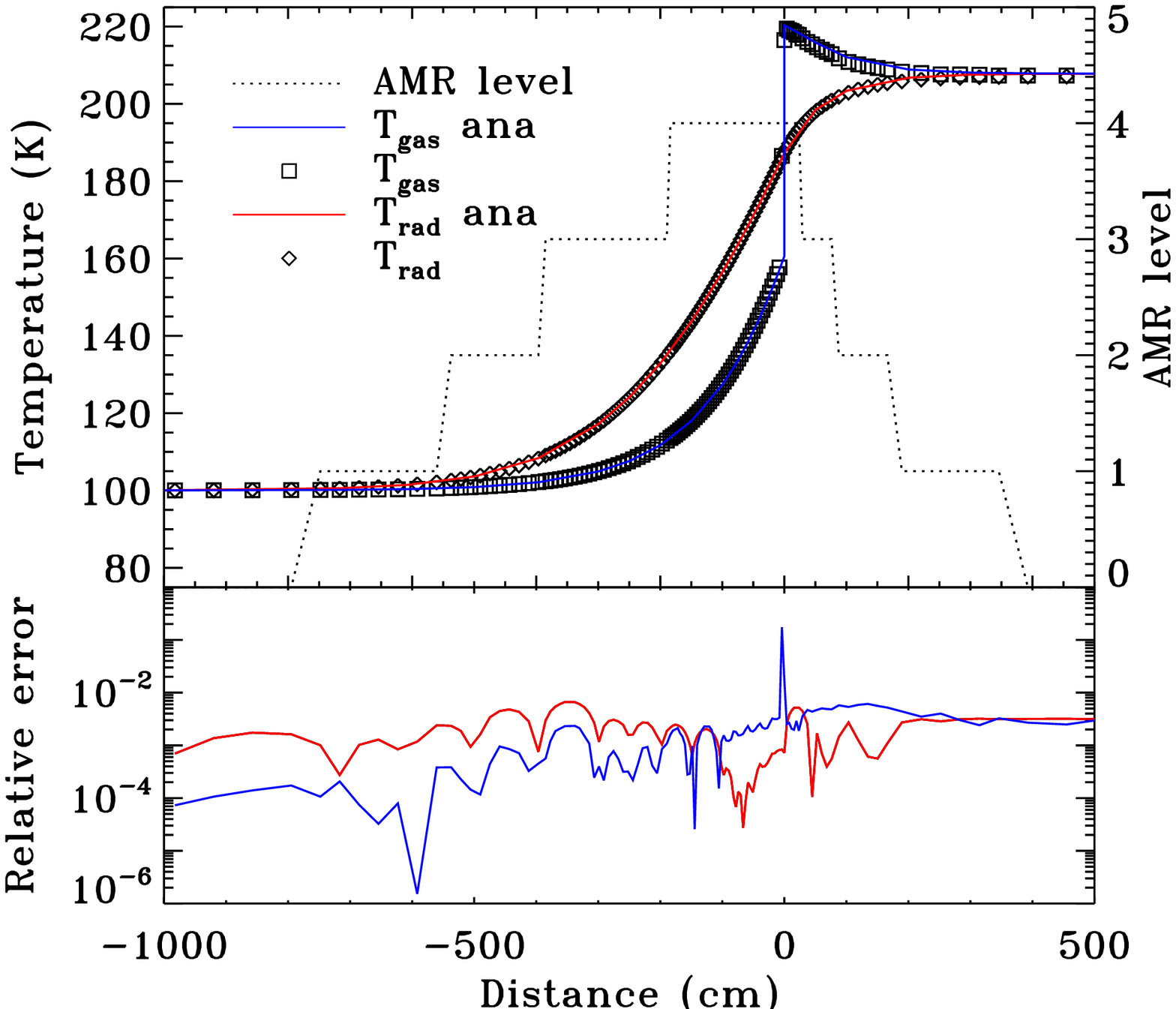}
\includegraphics[width=9.cm,height=7.65cm]{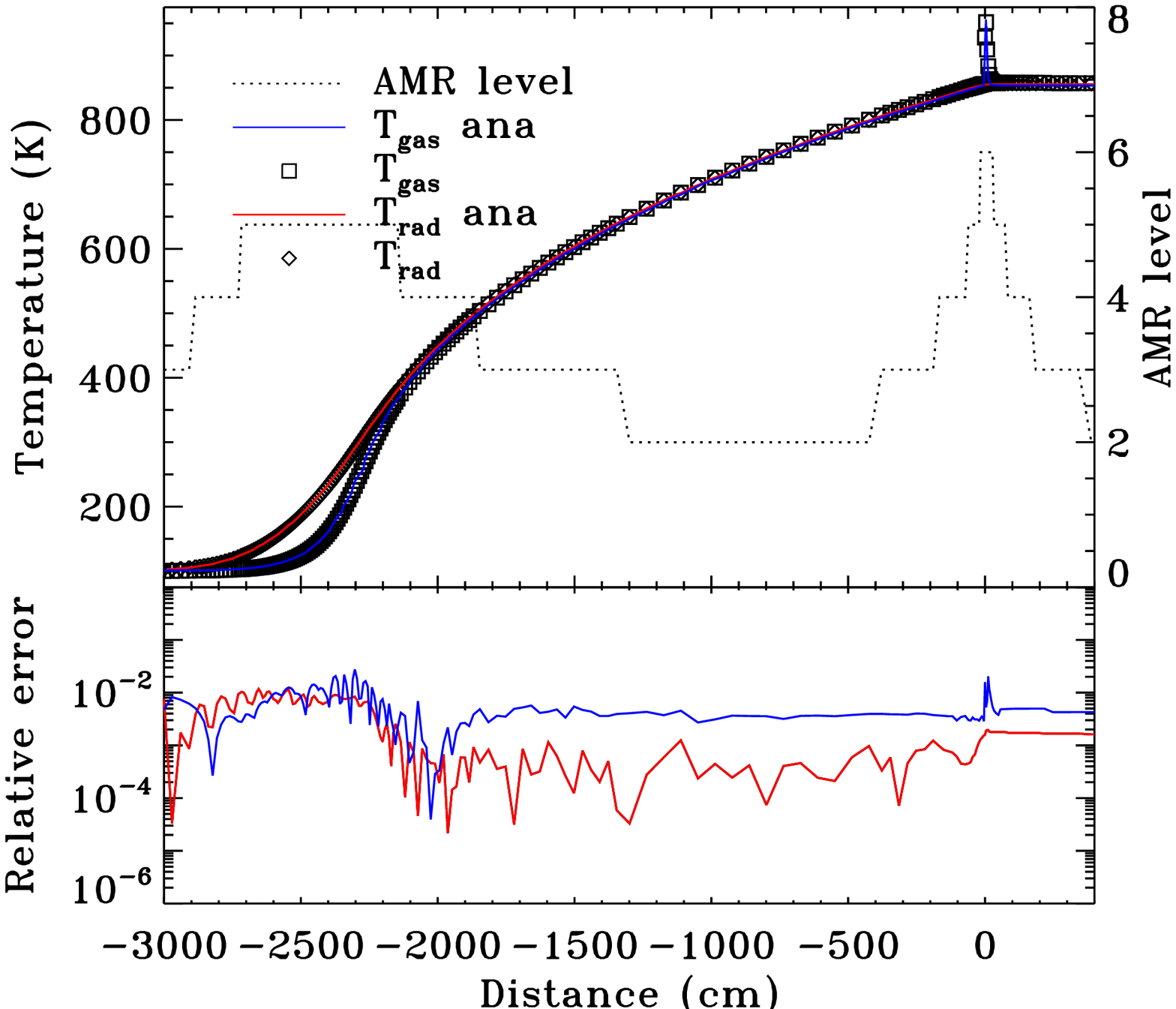}
\caption{Radiative shock test. {\it Left:} gas (blue line) and radiative (red line) temperature as a function of the distance to the shock  in the subcritical radiative shock ($M=2$) at time t=0.1 (top panel). The semi-analytic solutions of the gas (square) and radiative (diamond) temperatures from \cite{Lowrie_Edwards_2008} are over plotted. The AMR level is shown (right axis, dotted line). The bottom panel shows the corresponding relative error for the gas (blue) and radiative (red) temperatures.  {\it Right:} same as left panels for the supercritical radiative shock ($M=5$). }
\label{rshock}
\end{figure*}

In this second test, we check that the introduction of subcycling does not change the global second-order accuracy in space of the scheme. \cite{Commercon_2011} showed that the combination of AMR and a unique time step is globally second-order accurate in space. 

We consider a uniform density ($\rho=1$) within a box of L=1. We impose two different radiation energies at the domain boundaries, i.e., $E_\mathrm{r}(x=0)=4$ and $E_\mathrm{r}(x=1)=0.5$. The initial radiative energy is initialized as a step function, i.e., $E_{\mathrm{r},x<0.5}=4$ and $E_{\mathrm{r},x>0.5}=0.5$. The Rosseland opacity is a nonlinear function of $E_\mathrm{r}$, $\kappa_\mathrm{R}=10^{10}E_\mathrm{r}^{a}$ with $a=3/2$. The system relaxes towards a steady state solution so that we can test the accuracy of the scheme without any limitation due to truncation errors in time. The analytic stationary  solution $E_\mathrm{ana}(x)$ is given by 
\begin{equation}
E_\mathrm{ana}(x)= \left[\left(E_\mathrm{r}(x=1)^{a+1}-E_\mathrm{r}(x=0)^{a+1} \right)x +E_\mathrm{r}(x=0)^{a+1} \right]^{1/(n+1)}.
\end{equation}
The coarser grid comprises 8 cells, and we allow for up to four levels of refinement ($\ell_\mathrm{max}=4$, up to an effective resolution of 128 cells). The convergence criterion $\epsilon_\mathrm{conv}$ is set to $10^{-8}$ and the mesh is refined where  the radiative energy relative gradient exceeds 10 \%. We perform calculations using $\alpha=1$ and ATS (DTA model), the unique time step method of \cite{Commercon_2011}, DTU model, and for uniform grids ranging from 8 to 128 cells. In the DTA and DTU runs, we let vary the maximum level of refinement to achieve effective resolutions comparable to the uniform grid calculations.

Figure \ref{equilibrium} (top) shows the radiative energy profile for the DTA (black solid line) and DTU (red solid line) calculations when four levels of refinement are allowed and the analytic equilibrium solution (dashed line). The DTA and DTU calculations give nearly identical results, showing that the stationary state has been reached. Figure \ref{equilibrium} (bottom) shows the corresponding relative errors for the DTA and DTU models, and for a simulation run with a uniform grid of 256 cells. The relative errors is of the order of a few percents, except close to the right boundary, where the energy gradients are the strongest. 

Figure \ref{error_l2} shows the norm of the L$_2$ error, calculated as
\begin{equation}
\mathrm{L}_2 = \sqrt{\frac{1}{N_\mathrm{eff}}\sum_1^{N_\mathrm{cell}}{\left[\left(E_{\mathrm{r},i} - E_{\mathrm{ana}}(x_i) \right)\Delta x_i\right]^2}},
\end{equation} 
where $N_\mathrm{eff}$ is the number of cells given by the effective resolution at the maximum level of refinement and $N_\mathrm{cell}$ is the number of cells in the calculations. The L$_2$ norm is plotted as a function of the minimum grid size reached using AMR for the DTA (black square) and DTU (red triangle) models, and against the mesh size of the uniform grid models (blue cross). The first important result is that the scheme remains globally close to second-order accuracy in space even with AMR and ATS. The use of the AMR weakens the slope to first-order compared to the uniform grid one when more than two levels of refinement are used, since our numerical scheme is only first-order accurate in space at level interfaces. The ratio between the number of level interfaces and the total number of cells in the computational domain is high ($1/7$) and the errors are dominated by the one of the coarser level which explains that the second-order breaks. Second-order accuracy can be recovered when the coarse grid resolution is doubled \citep{Guillet_Teyssier_2011}. It is also worth mentioning that the error in the DTA model remains very close to the DTU one, which indicates that the error introduced by the lack of energy conservation at level boundaries is limited.  There are only 28 cells in the AMR DTA and DTU models for the most resolved calculations (128 cells effective resolution).

The calculations run with $\alpha=0$ crashed when we used more than two levels of refinement. This is due to negative energies that appear at the beginning of the calculations, when the mesh is refined only at the center and is thus in a situation similar to that of figure \ref{pb} close to the left box boundary. This shows the limitations of using Neumann boundary conditions at level interfaces. Last but not least, we see that the error only depends weakly on the method used to compute the flux at level boundaries, and the Dirichlet method, which does not conserve energy, gives remarkably good results.

\subsection{Radiative shocks}

\begin{figure*} [t]
\centering
\includegraphics[]{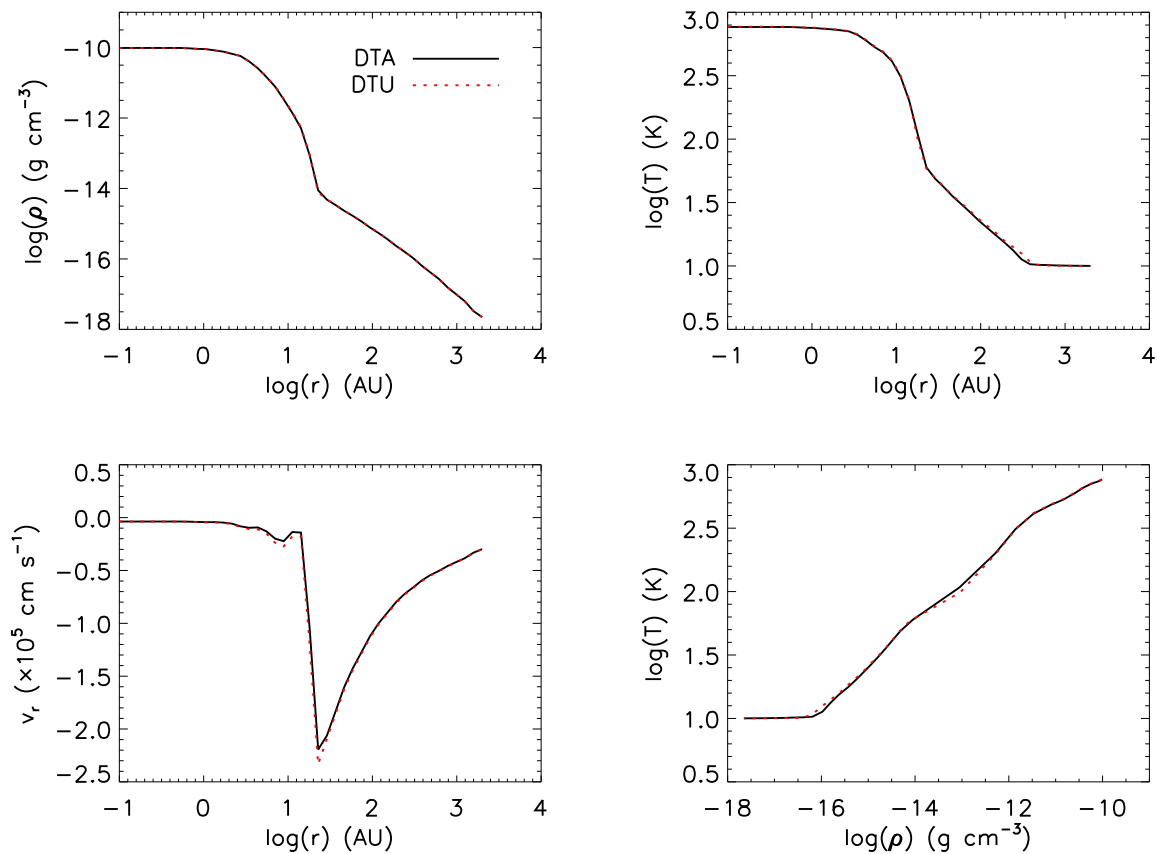}
\caption{Spherical collapse test. Density, temperature, and velocity profiles as a function of the radius and distribution of temperature as a function of the density for two calculations using adaptive time-stepping (DTA, solid black line) and with a unique time step (DTU, dotted red line) when the central density is $\rho_\mathrm{c}\sim 1\times10^{-10}$ g cm$^{-3}$.}
\label{collapse}
\end{figure*}

\begin{figure*} [t]
\centering
\includegraphics[]{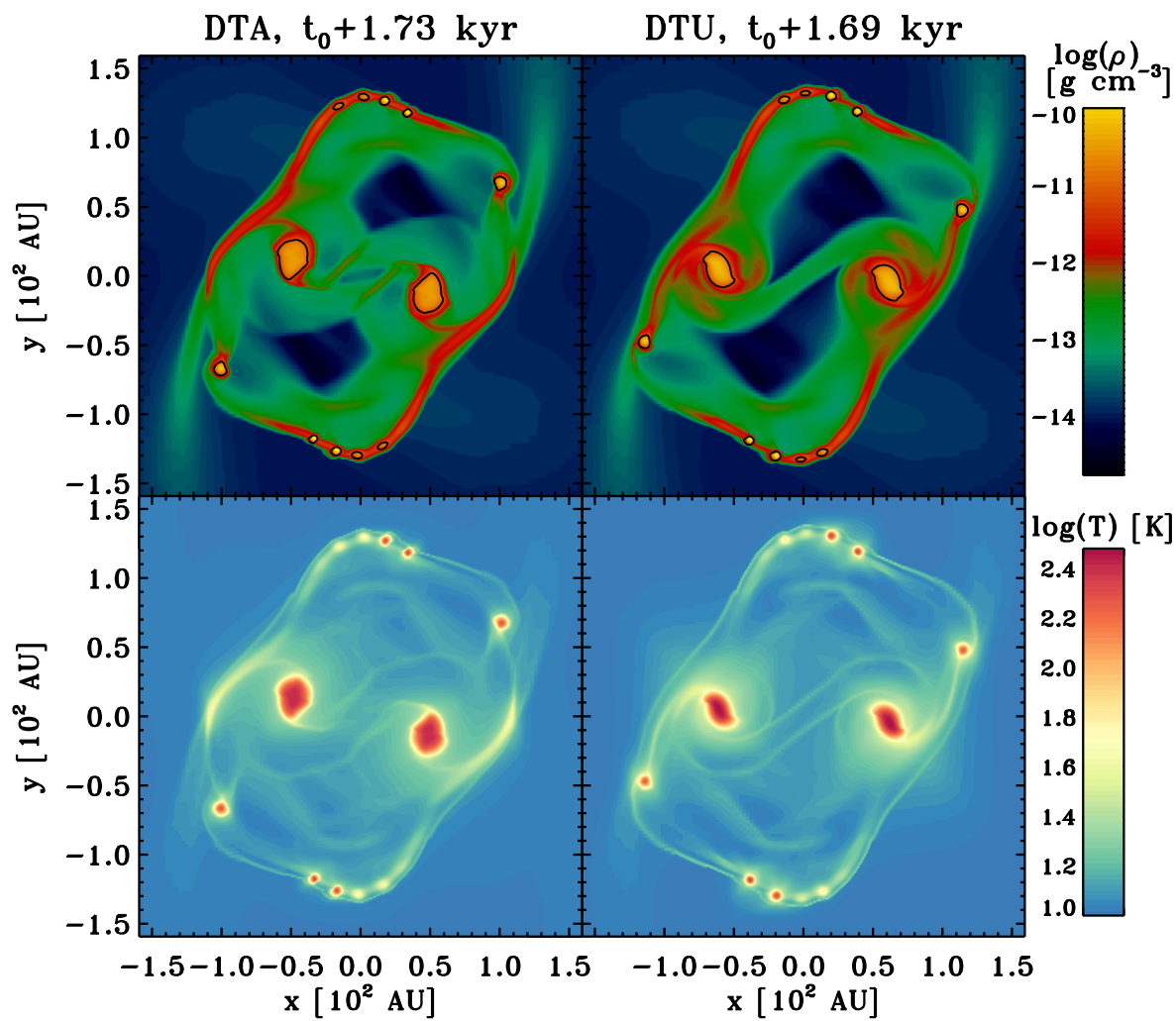}
\caption{Boss and Bodenheimer test. Density (top row) and gas temperature (bottom row) maps in the equatorial plane at time t$_0+1.73$ kyr for the DTA and t$_0+1.69$ kyr for the DTU (with t$_0$ the time at which the density first exceeds $10^{-13}$ g cm$^{-3}$). The black contours in the density maps shows the contour at $\rho=10^{12}$ cm$^{-3}$ ($\sim 3.87\times 10^{-12}$ g cm$^{-3}$) to identify the fragments.}
\label{BBtest}
\end{figure*}

\begin{figure*} [t]
\centering
\includegraphics[scale=0.8]{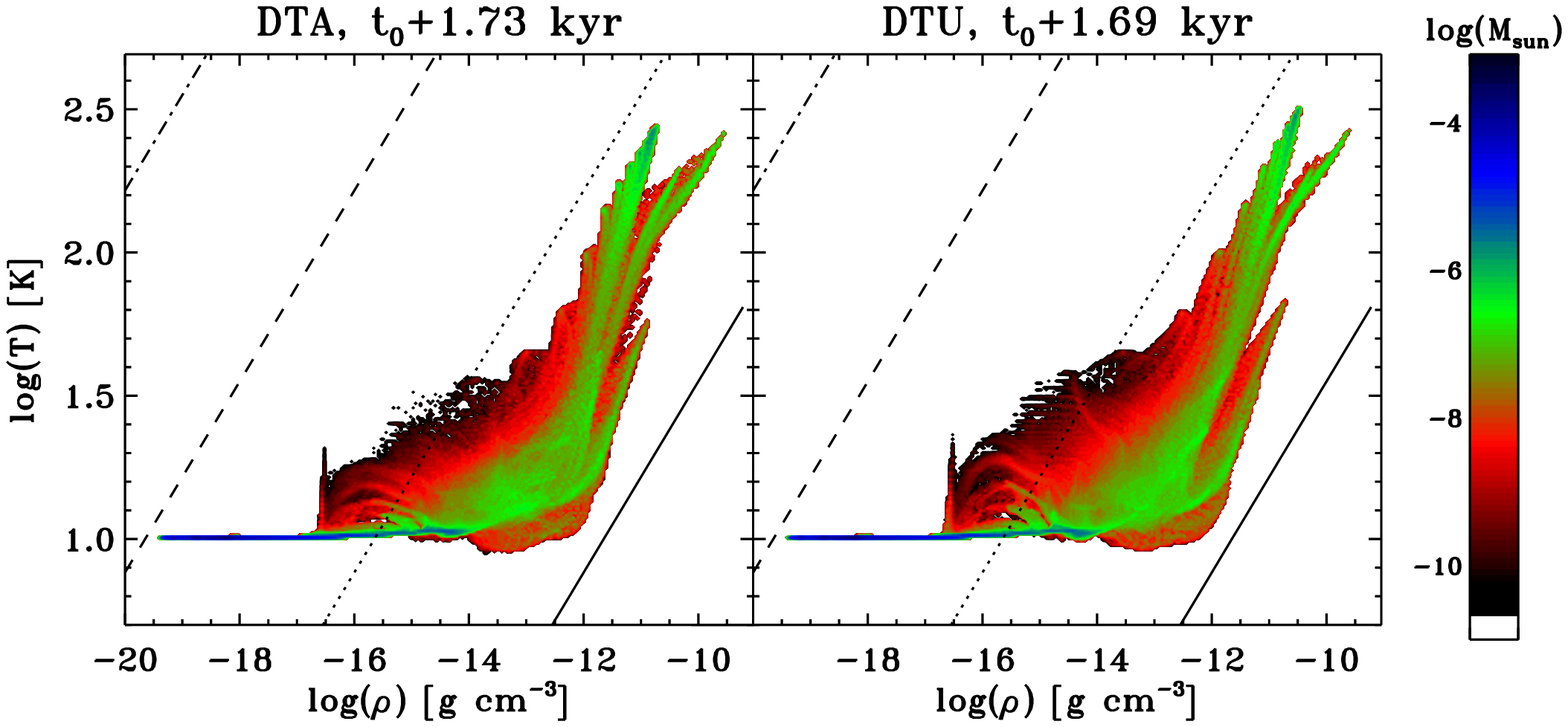}
\caption{Boss and Bodenheimer test. Distribution of the gas in the density-temperature plane for the DTA (left) and DTU (right) calculations and at the same time as in fig. \ref{BBtest}.  The color coding indicates the mass contained within each bin in the density-temperature plane. The oblique lines show the iso-Jeans mass, ranging from $10^{-5}$ M$_\odot$ to 10 M$_\odot$.}
\label{trho}
\end{figure*}

Radiative shocks are good laboratories to test our radiation hydrodynamics method. Classical analysis of radiative shocks can be found in \cite{Mihalas_book}. We choose initial conditions following \cite{Lowrie_Edwards_2008} who describe a semi-analytic method for the exact solution of radiative shock profiles with gray nonequilibrium diffusion in an optically thick medium. This setup has the advantage of resulting in a stationary shock and the semi-analytic solution can be directly compared with numerical results. We follow the initial setup of \cite{Zhang_et_al_2011} for the sub- and super-critical radiative shock corresponding to Mach numbers {\it M} of 2 and 5, respectively. 

The initial setup consists of a one dimensional region made of two uniform states which satisfy the jump relation for a radiating fluid in an optically thick medium \citep[e.g.,][]{Mihalas_book}. The boundary conditions are imposed at the initial state values throughout the calculation time.  We use an ideal gas equation of state, an adiabatic index $\gamma=5/3$, a mean molecular weight $\mu=1$, and an optically thick medium, i.e., $\lambda=1/3$. Matter and radiation are assumed to be initially in equilibrium, i.e., $T=T_\mathrm{r}$. The Planck and Rosseland opacities are set to $\kappa_\mathrm{P}=3.93 \times 10^{-5}$ cm$^{-1}$ and  $\kappa_\mathrm{R}=0.848902$ cm$^{-1}$. The initial grid is made of 32 cells and allows for maximum six additional levels of refinement. The refinement criteria are based on the gradients of the density  and the radiative energy. We use the $hll$ Riemann solver for the hydrodynamics with a CFL factor of $0.5$.  We use Dirichlet boundary conditions at level interfaces ($\alpha=0$). The convergence criterion $\epsilon_\mathrm{conv}$ is set to $10^{-8}$.

For the subcritical shock ($M=2$), the one dimensional region ranges from -1000 cm to 1000 cm and the discontinuity between the two initial states is located at $x=0$ cm. The two states characteristics are: $\rho_\mathrm{L}=5.45887\times10^{-13}$ g cm$^{-3}$, $u_\mathrm{L}=2.3545\times 10^5$ cm s$^{-1}$ and $T_\mathrm{L}=100$ K for the left state, and $\rho_\mathrm{R}=1.2479\times10^{-12}$ g cm$^{-3}$, $u_\mathrm{R}=1.03\times 10^5$ cm s$^{-1}$ and $T_\mathrm{R}=207.757$ K for the right state. A cell is refined when the relative gradient of density or radiative energy exceeds 5 \%, which provides   good resolution at the shock front and in the radiative precursor. Figure \ref{rshock} (left) shows the gas temperature (squares) and radiative temperature (diamonds) profiles (top panel), and the corresponding relative errors (bottom panel). The semi-analytic solutions are also plotted (blue line for the gas temperature and red line for the radiative temperature). We note that we needed to shift slightly the position $x=0$ since the shock moved a little from the initial $x=0$ discontinuity during a short period of adjustment to reach the steady state as the shock structure develops from the initial step profile. The AMR level (right axis) is plotted in dotted line. Once the steady state is reached, only four levels of refinement are used (130 cells in the AMR hierarchy), for an effective resolution of 512 cells ($3.9$ cm). The agreement between the numerical and the analytic solutions is very good as indicated in the relative error plots which shows errors below 1\% except at the location of the gas temperature spike. The shock is captured within only two cells thanks to the AMR capabilities. Compared to our previous unique time step method, the gain in CPU time is about a factor 50. 

For the supercritical shock ($M=5$), the one dimensional domain ranges from -4000 cm to 4000 cm and the discontinuity between the two initial states is placed at $x=0$ cm. The left state values are identical to those of the subcritical shock and the right state values read: $\rho_\mathrm{R}=1.964050\times10^{-12}$ g cm$^{-3}$, $u_\mathrm{R}=1.63\times 10^5$ cm s$^{-1}$ and $T_\mathrm{R}=855.72$ K. A cell is refined when the gradients of density and radiative energy exceed 1 \% and 10\% respectively. Figure \ref{rshock} (right) shows the profiles of radiative and gas temperatures (and the corresponding relative errors) following the same nomenclature as Fig. \ref{rshock} (left). In this case, the effective resolution reached at the shock location is $\sim 0.98$ cm (2048 cells) and the total number of cell in the computational domain is 225. The numerical and semi-analytic solutions agree again very well with relative errors of a few percent at most. Compared to the unique time step method, the gain in CPU time is about a factor $\sim 30$ in this case.

\section{Application to star formation}
Adaptive mesh-refinement is particularly well-suited for complex problems using deep hierarchies of levels such as those generated in collapse calculations. Moreover, in the star formation framework, radiative transfer plays a key role in the thermal behavior of the collapsing gas and alters dramatically the fragmentation of prestellar cores \citep[e.g.,][]{Bate_2009,Offner_2009,Commercon_2011_3}. The method we presented in Paper I has been used with success in star formation studies \citep{Commercon_2010L,Commercon_2012,Hincelin_2013} and successfully compared to 1D spherical calculations of \cite{Commercon_2011_2}. Nevertheless, this method was time-consuming because of the unique time stepping. The improvement we present in this paper with adaptive time-stepping is thus of prime importance for 
star formation purposes.

\subsection{Spherical collapse}

We present in this section a spherical collapse test of an isolated and gravitationally unstable 1 M$_\odot$ sphere of molecular gas, similar to the one  presented in Paper I. We wish to compare the results obtained with ATS and with a unique time step. We consider a uniform density ($\rho_0=1.14\times 10^{-18}$ g cm$^{-3}$) and temperature ($T_0=10$ K) dense core. The ratio between the thermal and gravitational energies, $\alpha_\mathrm{therm}=5R_0k_\mathrm{B}T_0/2G M_0\mu m_\mathrm{H}$, is 0.5 (free fall time $t_\mathrm{ff}\sim 62.3 $ kyr), which gives an initial radius $R_0\sim 0.024$ pc. The ratio of specific heats is $\gamma=5/3$ and the mean molecular weight is $\mu=2.33$. For the opacity, we use the gray 
 opacities of \cite{Semenov_et_al_2003A&A} as tabulated in \cite{Vaytet_2012} for homogeneous spherical dust grains and normal iron
content in the silicates (Fe/(Fe+Mg) = 0.3). The coarsest grid is made of $32^3$ cells and we use a refinement criterion based on the local Jeans length, which ensures that the latter is always resolved by at least eight cells. We use the \cite{Minerbo_1978JQSRT} flux limiter, the $hll$ Riemann solver, and a hydrodynamic CFL factor of $0.8$.  For the simulations with ATS, we use Dirichlet boundary conditions at level interfaces ($\alpha=0$). The convergence criterion $\epsilon_\mathrm{conv}$ is set to $10^{-4}$ for the iterative solver. We run the calculations until the late evolution of the first Larson core \citep{Larson_1969}. 

Figure \ref{collapse} shows the density, temperature, and velocity profiles as a function of the radius, and the corresponding distribution of temperature as a function of density for the calculations with ATS (DTA, solid black line) and with a unique time step (DTU, dotted red line) when the central density is $\rho_\mathrm{c}\sim 1\times10^{-10}$ g cm$^{-3}$. The two calculations agree remarkably well for all the different quantities given that the hydrodynamic and the radiation solvers are subcycled in the DTA calculations. We only see a few differences at the tail of the radiative precursor in the temperature profiles (more extended in the DTU model). The first core mass and radius are respectively $6.24\times 10^{-2}$ M$_\odot$ and 14.2 AU for the DTA and $6.37\times 10^{-2}$ M$_\odot$ and 14.2 AU for the DTU. The acceleration in term of CPU time thanks to ATS is about a factor 25 and the calculations have been run on eight processors.

\subsection{Boss \& Bodenheimer test}

This last test revisits a well-known numerical exercise on protostellar collapse. It is based on the early work by \cite{Boss_1979}. It consists of the collapse of a uniform density and temperature sphere in solid body rotation with an azimuthal density perturbation of amplitude $A$. This type of test invokes many physical processes: hydrodynamics, gravity, and radiative transfer. In addition, the high nonlinearity of the problem tends to shorten the horizon of predictability so that the comparison between two methods is challenging \citep[e.g.,][]{Commercon_2008}. 
We choose the same ratio of thermal to gravitational energy as in the last section, $\alpha_\mathrm{therm}=0.5$, a perturbation amplitude $A=0.1$, and the angular velocity $\Omega_0$ is set by the ratio of rotational to gravitational energy $\beta_\mathrm{rot}=R_0^3\Omega_0^2/3GM_0 =0.4$. The model has a high initial rotation, which favors the formation of a  large disk that is prone to fragmentation. We use the same initial and numerical parameters as in the previous test, except for the refinement criterion which was increased to ten points per Jeans length. The maximum effective resolution that is reached is 131072 cells ($2^{17}$, $\sim 0.15$ AU), corresponding to twelve additional levels of refinement. We ran two calculations, one with a unique time step (DTU) and another one with ATS (DTA). In the DTA calculations, the five first levels share the same time steps and the seven other use ATS.

Figure \ref{BBtest} shows the density and temperature maps for the two models, at time t$_0+1.73$ kyr for the DTA and t$_0+1.69$ kyr for the DTU (with t$_0$ the time at which the density first exceeds $10^{-13}$ g cm$^{-3}$). The qualitative agreement between the two calculations is good, given that in this comparison of methods, not only is the radiative solver subcycled, but also the hydrodynamics and the gravity solvers. The collapsing cores yield the same number of fragments (ten at this time). The mass within the fragments, i.e., where the density exceeds $10^{12}$ cm$^{-3}$, is 0.098 M$_\odot$ for DTA and 0.1 M$_\odot$ for DTU. The biggest fragments have a mass of $3.45\times10^{-2}$ M$_\odot$ for DTA and $3.56\times10^{-2}$ M$_\odot$ for DTU. The temperature maps show also similar features, such as temperature spikes in the shocked region and heated regions around the fragments. Figure \ref{trho} shows the density-temperature distribution in the two calculations at the same time as in fig. \ref{BBtest}. The color coding indicates the mass contained within each bin in the density-temperature distribution. Again, the agreement between the two methods is  good, in particular for the green area that represents most of the mass contained within or around the fragments. The typical Jeans mass of the fragments are also similar. In this last test, the gain in term of CPU time is about a factor 5.

\section{Conclusion and future work}

We have presented in this paper a new method for implicit solvers on adaptive mesh-refinement grids in the context of diffusion problems. The method can deal with an adaptive time-stepping strategy such as those used by hydrodynamical solvers. Our method has been successfully implemented in the {\ttfamily RAMSES} code for radiation-hydrodynamics using the flux-limited diffusion approximation. The principle of this new solver is to consider each level of the AMR hierarchy independently from the others and to use simple recipes for the boundary conditions at level interfaces (Dirichlet, Neumann, and Robin conditions). We have demonstrated that each of the different boundary conditions has its pros and cons. In particular, the Neumann conditions, which ensures energy conservation, can lead to negative energies and errors in the deposit of energy that is stored at level interfaces. On the opposite, we showed that the simple Dirichlet condition is much more robust even if it does not strictly conserve energy. We tested our method against classical numerical exercises (diffusion test, radiative shocks) and compared our results with analytic solutions. The new method is close to second-order accuracy in space and the error only depends weakly on the type of boundary condition used at level interfaces. We applied the method to a star formation test case and successfully compared  the new results to the ones obtained using the unique time step method presented in Paper I. The gain in CPU time can vary from a factor 5 to a factor 50, depending on the problem.

This new method makes use of a simple conjugate gradient algorithm as an iterative solver to integrate the diffusion operator. Since all the levels evolve independently from the others, we plan to allow the use of super-time-stepping \citep{Alexiades_1996,Commercon_2011} and of explicit time integration depending on the ratio between the Courant condition for the diffusion and the one for the hydrodynamics. Concerning radiative transfer, the method presented in this paper is limited to gray radiation. Extension towards multigroup radiative transfer is in progress \citep[e.g.,][]{Vaytet_2012}.

Last but not least, the implicit adaptive time-stepping can be applied to the study of astrophysical structures in which other diffusion-like problems such as the propagation of cosmic rays and the anisotropic electronic conduction are involved.

\begin{acknowledgements}
We thank the anonymous referee for his/her comments that improve the quality of the paper. BC acknowledges Neil Vaytet and Patrick Hennebelle for useful comments and discussions. The work presented has been supported by the French ANR Retour Postdoc program.
\end{acknowledgements}

\bibliographystyle{aa}
\bibliography{biblio2}

\begin{thebibliography}{26}
\expandafter\ifx\csname natexlab\endcsname\relax\def\natexlab#1{#1}\fi

\bibitem[{Alexiades {et~al.}(1996)Alexiades, Amiez, \&
  Gremaud}]{Alexiades_1996}
Alexiades, V., Amiez, G., \& Gremaud, P.-A. 1996, Com. Num. Meth. Eng, 12, 12

\bibitem[{{Almgren} {et~al.}(2010){Almgren}, {Beckner}, {Bell}, {Day},
  {Howell}, {Joggerst}, {Lijewski}, {Nonaka}, {Singer}, \&
  {Zingale}}]{Almgren_2010}
{Almgren}, A.~S., {Beckner}, V.~E., {Bell}, J.~B., {et~al.} 2010, \apj, 715,
  1221

\bibitem[{{Bate}(2009)}]{Bate_2009}
{Bate}, M.~R. 2009, \mnras, 392, 1363

\bibitem[{{Boss} \& {Bodenheimer}(1979)}]{Boss_1979}
{Boss}, A.~P. \& {Bodenheimer}, P. 1979, \apj, 234, 289

\bibitem[{{Commer{\c c}on} {et~al.}(2011{\natexlab{a}}){Commer{\c c}on},
  {Audit}, {Chabrier}, \& {Chi{\`e}ze}}]{Commercon_2011_2}
{Commer{\c c}on}, B., {Audit}, E., {Chabrier}, G., \& {Chi{\`e}ze}, J.-P.
  2011{\natexlab{a}}, \aap, 530, A13

\bibitem[{{Commer{\c c}on} {et~al.}(2008){Commer{\c c}on}, {Hennebelle},
  {Audit}, {Chabrier}, \& {Teyssier}}]{Commercon_2008}
{Commer{\c c}on}, B., {Hennebelle}, P., {Audit}, E., {Chabrier}, G., \&
  {Teyssier}, R. 2008, \aap, 482, 371

\bibitem[{{Commer{\c c}on} {et~al.}(2010){Commer{\c c}on}, {Hennebelle},
  {Audit}, {Chabrier}, \& {Teyssier}}]{Commercon_2010L}
{Commer{\c c}on}, B., {Hennebelle}, P., {Audit}, E., {Chabrier}, G., \&
  {Teyssier}, R. 2010, \aap, 510, L3+

\bibitem[{{Commer{\c c}on} {et~al.}(2011{\natexlab{b}}){Commer{\c c}on},
  {Hennebelle}, \& {Henning}}]{Commercon_2011_3}
{Commer{\c c}on}, B., {Hennebelle}, P., \& {Henning}, T. 2011{\natexlab{b}},
  \apjl, 742, L9

\bibitem[{{Commer{\c c}on} {et~al.}(2012){Commer{\c c}on}, {Launhardt},
  {Dullemond}, \& {Henning}}]{Commercon_2012}
{Commer{\c c}on}, B., {Launhardt}, R., {Dullemond}, C., \& {Henning}, T. 2012,
  \aap, 545, A98

\bibitem[{{Commer{\c c}on} {et~al.}(2011{\natexlab{c}}){Commer{\c c}on},
  {Teyssier}, {Audit}, {Hennebelle}, \& {Chabrier}}]{Commercon_2011}
{Commer{\c c}on}, B., {Teyssier}, R., {Audit}, E., {Hennebelle}, P., \&
  {Chabrier}, G. 2011{\natexlab{c}}, \aap, 529, A35

\bibitem[{{Guillet} \& {Teyssier}(2011)}]{Guillet_Teyssier_2011}
{Guillet}, T. \& {Teyssier}, R. 2011, Journal of Computational Physics, 230,
  4756

\bibitem[{{Hincelin} {et~al.}(2013){Hincelin}, {Wakelam}, {Commer{\c c}on},
  {Hersant}, \& {Guilloteau}}]{Hincelin_2013}
{Hincelin}, U., {Wakelam}, V., {Commer{\c c}on}, B., {Hersant}, F., \&
  {Guilloteau}, S. 2013, \apj, 775, 44

\bibitem[{{Howell} \& {Greenough}(2003)}]{Howell_Greenough_03}
{Howell}, L.~H. \& {Greenough}, J.~A. 2003, Journal of Computational Physics,
  184, 53

\bibitem[{{Larson}(1969)}]{Larson_1969}
{Larson}, R.~B. 1969, \mnras, 145, 271

\bibitem[{{Levermore} \& {Pomraning}(1981)}]{Levermore_Pomraning_1981ApJ}
{Levermore}, C.~D. \& {Pomraning}, G.~C. 1981, \apj, 248, 321

\bibitem[{{Lowrie} \& {Edwards}(2008)}]{Lowrie_Edwards_2008}
{Lowrie}, R.~B. \& {Edwards}, J.~D. 2008, Shock Waves, 18, 129

\bibitem[{{Mihalas} \& {Mihalas}(1984)}]{Mihalas_book}
{Mihalas}, D. \& {Mihalas}, B.~W. 1984, {Foundations of radiation
  hydrodynamics}, ed. D.~{Mihalas} \& B.~W. {Mihalas}

\bibitem[{{Minerbo}(1978)}]{Minerbo_1978JQSRT}
{Minerbo}, G.~N. 1978, \jqsrt, 20, 541

\bibitem[{{Offner} {et~al.}(2009){Offner}, {Klein}, {McKee}, \&
  {Krumholz}}]{Offner_2009}
{Offner}, S.~S.~R., {Klein}, R.~I., {McKee}, C.~F., \& {Krumholz}, M.~R. 2009,
  \apj, 703, 131

\bibitem[{{Semenov} {et~al.}(2003){Semenov}, {Henning}, {Helling}, {Ilgner}, \&
  {Sedlmayr}}]{Semenov_et_al_2003A&A}
{Semenov}, D., {Henning}, T., {Helling}, C., {Ilgner}, M., \& {Sedlmayr}, E.
  2003, \aap, 410, 611

\bibitem[{Tang(1992)}]{Tang_1992}
Tang, W.~P. 1992, SIAM J. Sci. Stat. Comput., 13, 573

\bibitem[{{Teyssier}(2002)}]{teyssier-2002}
{Teyssier}, R. 2002, \aap, 385, 337

\bibitem[{{The Enzo Collaboration} {et~al.}(2013){The Enzo Collaboration},
  {Bryan}, {Norman}, {O'Shea}, {Abel}, {Wise}, {Turk}, {Reynolds}, {Collins},
  {Wang}, {Skillman}, {Smith}, {Harkness}, {Bordner}, {Kim}, {Kuhlen}, {Xu},
  {Goldbaum}, {Hummels}, {Kritsuk}, {Tasker}, {Skory}, {Simpson}, {Hahn},
  {Oishi}, {So}, {Zhao}, {Cen}, \& {Li}}]{Enzo_2013}
{The Enzo Collaboration}, {Bryan}, G.~L., {Norman}, M.~L., {et~al.} 2013, ArXiv
  e-prints

\bibitem[{{Turner} \& {Stone}(2001)}]{Turner_stone_01}
{Turner}, N.~J. \& {Stone}, J.~M. 2001, \apjs, 135, 95

\bibitem[{{Vaytet} {et~al.}(2012){Vaytet}, {Audit}, {Chabrier}, {Commer{\c
  c}on}, \& {Masson}}]{Vaytet_2012}
{Vaytet}, N., {Audit}, E., {Chabrier}, G., {Commer{\c c}on}, B., \& {Masson},
  J. 2012, \aap, 543, A60

\bibitem[{{Zhang} {et~al.}(2011){Zhang}, {Howell}, {Almgren}, {Burrows}, \&
  {Bell}}]{Zhang_et_al_2011}
{Zhang}, W., {Howell}, L., {Almgren}, A., {Burrows}, A., \& {Bell}, J. 2011,
  \apjs, 196, 20

\end{thebibliography}

\end{document}